\documentclass[]{scrartcl}

\usepackage{amsmath,amssymb}
\usepackage{graphicx}
\usepackage{cite}

\usepackage{color} 
\usepackage{geometry}
\title{Dynamic network and epigenetic landscape model of a regulatory core underlying spontaneous immortalization and epithelial carcinogenesis}

\author{M\'{e}ndez-L\'{o}pez LF\textsuperscript{1,2,\dag}, Davila-Velderrain J\textsuperscript{2,\dag}, Enr\'iquez-Olgu\'in C\textsuperscript{3} \\ Dom\'inguez-H\"{u}ttinger E\textsuperscript{1,2}, Martinez-Garcia JC\textsuperscript{3*}, Alvarez-Buylla ER\textsuperscript{1,2,*}}

\date{}

\begin{document}

\maketitle

{\raggedright
	
	{\footnotesize	
		\bf{1} Instituto de Ecolog\'ia, Universidad Nacional Aut\'onoma de M\'exico, Cd. Universitaria, M\'exico, D.F. 04510, M\'exico \\
		\bf{2} Centro de Ciencias de la Complejidad (C3), Universidad Nacional Aut\'onoma de M\'exico, Cd. Universitaria, M\'exico, D.F. 04510, M\'exico \\
		\bf{3} Departamento de Control Autom\'atico, Instituto Polit\'ecnico Nacional, A. P. 14-740, 07300 M\'exico, DF, M\'exico
	}
}

$\ast$ Corresponding authors: juancarlos\_martinez-garcia@conciliencia.org, eabuylla@gmail.com \\
$\dag$ These authors contributed equally to this work

\begin{abstract}
Tumorigenic transformation of human epithelial cells {\em in vitro} has been described experimentally as the potential result of a process known as {\em spontaneous immortalization}. In this process a generic series of cell--state transitions occur in which normal epithelial cells acquire a senescent state, later surpassed to attain first a mesenchymal state and then a final mesenchymal stem--like phenotype, with potential tumorigenic behavior. In this paper we integrate published data on the molecular components and interactions that have been described as key regulators of such cell states and transitions. A large network, that is provided, is constructed and then reduced with the aim of recovering a minimal regulatory core incorporating the necessary and sufficient restrictions to recover the observed cell states and their generic progression patterns in epithelial--mesenchymal transition. Data is formalized into logical regulatory rules that govern the dynamics of each of the network's components as a function of the states of its regulators. The proposed core gene regulatory network attains only three steady--state gene expression configurations that correspond to the profiles characteristic of normal epithelial, senescent, and mesenchymal stem--like cells. Interestingly, epigenetic landscape analyses of the uncovered network shows that it also recovers the generic time--ordered transitions documented during tumorigenic transformation {\em in vitro} of epithelial cells. The latter strongly correlate with the patterns observed during the progressive pathological description of epithelial carcinogenesis {\em in vivo}.
\end{abstract}

\section*{Introduction}

Nearly 84\% of cancers diagnosed in human adults are carcinomas (i.e., cancer of epithelial origin), and their emergence is strongly associated with both an underlying chronic inflammatory process and with aging \cite{anand2008cancer}. The precise role and the contribution of these two processes to the origin, progression, and detected clinic behavior of epithelial cancers remains elusive, however. The current  general assumption is that aging and inflammation increase the chance of accumulating somatic mutations, and this genetic instability ultimately leads to carcinoma. However, this view does not offer a logical or mechanistic explanation for well--documented observations. For example: (1) cancer cells show morphological and transcriptional convergences despite their diverse origin, (2) carcinogenesis recapitulates embryonic processes, (3) cancer behavior can be acquired in the absence of mutations through trans-- or dedifferentiation, and (4) cancer cells can be ``normalized" by several experimental means \cite{huang2011intrinsic, mani2008epithelial, huang2009non, ben2008embryonic}. Moreover, it is well--known that different carcinomas share the same cellular processes and histological stages or progression patterns, as well as robust associations with lifestyle factors  \cite{kelloff2007assessing}. These empirical observations suggests that, in analogy to normal development, the human genome is associated with an underlying robust mechanism restricting cell states and temporal progression patterns that are characteristic of epithelial carcinogenesis. In accordance with this view, other researchers have previously proposed that cancer can be considered a developmental disease \cite{virchow1860cellular, huang2009cancer}. \\

In systems biology it is common to understand both cell differentiation and development in terms of dynamical systems theory. In this framework, the genome of a cell is directly mapped into a global and multi--stable gene regulatory network (GRN) whose dynamics yields several (quasi)stationary and stable distinct phenotypic cellular states  \cite{mendoza1998gene, espinosa2004gene, huang2009complex, alvarez2010abc, huang2009reprogramming, kaneko2011characterization}. That is, the same genome robustly generates multiple discrete cellular phenotypes through developmental dynamics \cite{alvarez2010abc,huang2012molecular, davila2014bridging}. These stable phenotypic states are called {\em attractors} and correspond to configurations of gene or protein activation states that underlie the cellular fates or phenotypes -- {\em i.e.}, which thus constitute biological observables. Therefore, developmental processes -- cellular differentiation events in particular -- are formalized in temporal terms as attractor's ({\em i.e.}, cell states) transitions. Here we adopt such approach to study the cell states attained and the time--ordered transitions observed during the tumorigenic transformation of epithelial cells cultured {\em in vitro} that surpass a senescent state; a process known as {\em spontaneous immortalization}. \\

Experimental findings in molecular and cell biology of cancer research have revealed that it is possible to recover cells with cancer--like phenotypes through some specific cellular transitions. This has been shown particularly in carcinomas \cite{mani2008epithelial, xu2009tgf, battula2010epithelial, li2012epithelial}. By a cellular transition we refer to  a differentiation event in which a certain cell acquires a discretely different cellular phenotype. For example, the process called epithelial--mesenchymal transition (EMT) comprises a stereotypical cell state transition in which epithelial cells exposed, for example,  to cytokines, are induced to undergo a discrete phenotypic change acquiring a mesenchymal phenotype \cite{li2012epithelial, xu2009tgf}. Interestingly, through inflammation--induced EMT epithelial cells surpass senescence, and undergo spontaneous immortalization.  Cells that emerge from this process manifest mesenchymal stem--like properties and are capable of developing cancer in murine models \cite{mani2008epithelial, battula2010epithelial}. Furthermore, these cells are difficult to distinguish phenotypically and in terms of the transcription factors that they express from either the so--called cancer stem cells (also known as tumor initiating cells) or from embryonic stem cells \cite{morel2008generation, neph2012circuitry}. \\

In the present work, we hypothesize that a generic series of cell state transitions widely observed and robustly induced by inflammation in cell cultures during spontaneous immortalization naturally result from the self--organized behavior emerging from an underlying intracellular GRN. During this process, normal epithelial cells first acquire a senescent state, to finally attain a mesenchymal stem--like cellular state with a potential tumorigenic behavior. We speculate that tissue--level conditions associated with a bad prognosis, such as a pro--inflammatory milieu, may increase the rate of occurrence of these same transitions {\em in vivo} promoting as a result the emergence and progression of epithelial cancer. \\  

In an attempt to provide mechanistic insights into the regulation of the aforementioned observed cell--fates specification, as well as the time--ordered cell--state transitions, we propose here a cellular level GRN model that integrates the available experimental data concerning the main molecular components and interactions related to the emergence and progression of carcinomas. We propose a large GRN of 41 nodes that integrates cellular processes thoroughly studied experimentally, but which have not been integrated before into a single GRN. Specifically, the large GRN model includes key molecular regulators that: (1) characterize the cellular phenotypes of epithelial, mesenchymal, and senescent cells; (2) are involved in the induction of the cellular processes of replicative senescence, cellular inflammation, and EMT; and (3) characterize the phenotypic changes undergone by cells emerging from these processes ({\em i.e.} mesenchymal stem--like cells). To obtain a minimal regulatory core for further dynamical analyses we formally reduced the large GRN. We show that the proposed regulatory core module displays an orchestrating robust behavior akin to that seen in other developmental regulatory modules previously characterized with similar formal approaches (see, for example \cite{mendoza1998gene, espinosa2004gene, azpeitia2010single, azpeitia2014gene}). Specifically, by proposing logical functions grounded on experimental data for this regulatory core module and by analyzing its behavior following conventional Boolean GRN dynamical approaches, we show that the uncovered minimal GRN converges only to three attractors. The uncovered states correspond to the expected gene expression configurations that have been observed for normal epithelial, senescent and stem--like mesenchymal cellular fates. Additionally, we also explore the GRN Epigenetic Landscape using a stochastic version of the model (following: \cite{alvarez2008floral, davila2015modeling}) in order to address if the proposed GRN also restricts or underlies the generic temporal sequence with which cell states occur in cell cultures and which correlate with observed patterns of cell--type enrichment during pathological descriptions of carcinoma progression.

\section*{Results}

\subsection*{Gene Regulatory Network Construction}

Following a bottom--up and an expert knowledge approach we propose a set of cellular dynamical processes ubiquitous to epithelial carcinogenesis, namely: replicative cellular senescence, inflammation, and  epithelial--mesenchymal transition (EMT). The cellular phenotypes epithelial, senescent, and mesenchymal cell--types -- as well as a mesenchymal embryonic--like state; have been largely characterized as biological observables involved in such processes.
We provide further definitions of these -- and associated -- phenotypes and processes in our complementary Text S1. We take this information as a methodological basis to integrate a generic dynamical network model of epithelial carcinogenesis. As a first step in network integration, based on an extensive literature search (see Methods and Text S1), we assembled a set of transcription factors (TFs) and additional molecules involved in the establishment and regulation of these cellular states and processes. Subsequently, we manually retrieved documented regulatory interactions among the molecules, considering only those supported by experimental evidence. For a detailed description of the published information for each interaction proposed see Text S1. The constructed large GRN is shown in Figure 1 (see Methods). TFs are represented in graphical terms by squares and the rest of the molecules by circles. The identified large network consists of 41 nodes and 97 interactions; it includes 12 TFs which can be considered as key regulators of the processes under consideration. Colors indicate the association that each node hold with specific cellular phenotypes or processes being considered: epithelial (green), mesenchymal (orange), inflammation (red), senescence and DNA damage (blue), cell--cycle (purple), and polycomb complex (yellow). 

\subsection*{The Proposed Network is Enriched with Cancer Pathways}

In order to provide additional partial support for the association of the bio-molecular set of regulatory interactions that we have manually curated based on published data with the processes under consideration, as well as with carcinoma, we performed a network--based gene set enrichment analysis (GSEA) (see Methods). Among the 13 pathways or processes reported as significant when taking the KEGG database as a reference, 9 (69\%) correspond to cancer pathways, namely: {\em Bladder cancer, Chronic myeloid leukemia, Pancreatic cancer, Glioma, Non--small cell lung cancer, Melanoma, Small cell lung cancer, Prostate cancer, and Thyroid cancer} -- note that 6 (66.6\%) of these correspond to carcinomas. On the other hand, when taking the GO Biological Process database as reference, among the significant results we found: {\em replicative senescence, cellular senescence, cell aging, activation of NF--$\kappa$B--inducing kinase activity, determination of adult life span, epithelial cell differentiation, and positive regulation of NF--$\kappa$B transcription factor activity} (see Table 1). Using network topological gene set analysis (see Methods) we found that, in addition to pathway enrichment, the topological signature of the molecules in the proposed network also shows a topological signature that is similar to the one shown by reference cancer pathways included in the KEGG database (see Figure S1). These results provide partial support for the proposed molecular players: given the current state of knowledge according to annotated databases, the set of molecules manually included in the proposed large network seems to be representative of the cellular phenotypes and processes considered as prior biological knowledge in our model. In addition, the molecular components included in the proposed large network are tightly associated with reference pathways of epithelial cancers.  

\subsection*{A Core Regulatory Network Module Underlying Spontaneous Immortalization}

We performed a knowledge--based network reduction of the large GRN in Figure 1 in order to derive a smaller, core GRN module for which both a topology and architecture with fully defined logical functions could be established, and which could also be analyzed as a dynamical system (see Methods). In addition, such regulatory core should comprise the necessary and sufficient set of nodes and interactions that integrate the processes involved in the large network and that could explain, at least in part, the restricted set of the cell--states and time--ordered transitions among them during spontaneous immortalization and epithelial cancer emergence/progression. We were able to define a set of molecular species  whose regulatory hierarchy, activity, and expression define the identity of the phenotypes of epithelial, mesenchymal, and senescent cells. We also converged to, and included, main regulators of replicative cellular senescence, inflammation--induced EMT, and determinants of an induced mesenchymal stem--like phenotype. Hence, after reduction we obtained a core GRN consisting of only 9 nodes: {\em ESE--2, Snai2, NF--$\kappa$B, E2F, p53, p16, Rb, Cyclin, and Telomerase}. Figure 1b shows the proposed core regulatory module (colored nodes) in the context of the larger proposed network. For details on how these 9 nodes were selected over the rest of the nodes see Text S1. In what follows we present a brief description of the nodes included in the reduced GRN, as well as some of the key molecular mechanism encoded in the regulatory logic. Although many of the nodes that are included in this regulatory core module have been thoroughly studied experimentally and in terms of their involvement in different types of cancer, the architecture and topology of the proposed regulatory core module is novel.           

\begin{description}
	\item[ESE--2] represents the activity of the TFs ESE--1, ESE--2, and ESE--3 (also known as ESX, E74--like factor 5, and EHF; respectively) -- for a table with synonyms Table E1 in supplementary file. These proteins belong to the subgroup ESE ({\em i.e.} epithelium--specific) of the TF family ETS. ESE--2 promotes its own expression and the expression of the other ESE TFs \cite{zhou1998novel, ma2003gene, escamilla2010genome}. On the other hand, ESE--2 represses Snai2 -- one of the main EMT promoting TFs -- expression by direct interaction with its promoter region \cite{chakrabarti2012elf5}.
	
	\item[p16] represents the activity of the INK4b--ARF--INK4a locus, which encodes for the proteins p16 and p14. Cellular senescence is molecularly characterized by the expression of the proteins p16 and p53 \cite{vernier2011regulation}. p16 indirectly inhibits E2F by inhibiting cyclins CDK 2,4 and 6, which in turn inhibit Rb \cite{mcconnell1999induced, villacanas2002structural}. On the other hand, the INK4b--ARF--INK4a -- and thus p16 -- is regulated by the activity of Polycomb--group proteins by means of promoter hypermethylation \cite{bracken2007polycomb}.
	
	\item[p53] represents the protein with the same name. The shortage of telomeric DNA seems to be recognized as DNA damage promoting  the activation of p53. In senescence, the activity of p16 and p53 over Rb, E2F and Cyclins invariably arrests the cell--cycle in the phase G1/G2 \cite{fang1999p21waf1,mao2012replicatively}. 
	
	\item[Rb] represents the cell--cycle regulator with the same name. Rb prevents cycle progression by forming a complex with the TF E2F \cite{chellappan1991e2f}.  
	
	\item[E2F] represents the TF with the same name. E2F regulates critical genes for adequate cell--cycle progression.
	
	\item[Cyclin] represents the activity of the complex Cyclin--dependent kinases (CDKs) known to inactivate Rb by phosphorylation. The latter, in turn, promotes the activity of E2F and cell--cycle progression \cite{byeon1998tumor}.
	
	\item[NF-$\kappa$B] represents cellular inflammation by the activity of the TF NF--$\kappa$B. Accordingly, with this node we also represent the effect of the cytokines transforming growth factor--beta (TGF--$\beta$), interleukin--6 (IL--6), and tumor necrosis factor alpha (TNF--$\alpha$). These three factors converge in the activation of NF--$\kappa$B by phosphorylating the inhibitor I$\kappa$B \cite{beausejour2003reversal, freudlsperger2012tgf}.
	
	\item[TELasa] represents the enzyme telomerase. This enzyme is responsible for the {\em de novo} synthesis of telomeres. Most human cell--types do not express telomerase; however, it is expressed on immortalized epithelial cells, and it is thought to be responsible for telomere extension in tumors \cite{harley1990telomeres}.
	
	\item[Snai2] this node includes the activity of the main TFs known to be directly associated with EMT regulation, namely:  Snai2 (Slug), Snail, Twist1, Twist2, ZEB1, ZEB2, and FOXC2. These TFs repress (induce) the expression of genes specific to epithelial (mesenchymal) cells \cite{mani2007mesenchyme, zeisberg2009biomarkers}. It has been proposed that there is a regulatory hierarchy driving EMT in which Snail activates Snai2, Twist, Zeb, and FOXC2. The latter, in turn, regulates Snail and Snai2 in a positive manner \cite{bolos2003transcription, mani2007mesenchyme, dave2011functional, casas2011snail2}. Regardless of a hierarchical interpretation, it is well--documented that these TFs maintain the mesenchymal phenotype in a coordinated fashion, showing co--expression patterns and regulatory crosstalk \cite{dave2011functional, casas2011snail2}. It has been suggested that among these TFs, Snai2 may be the strongest suppressor of the epithelial phenotype \cite{hajra2002slug}. However, we decided to represent the collective regulatory activity of the mesenchymal TFs using Snai2 based on the recent experimental demonstration of an antagonistic relation between Snai2 and ESE--2. Specifically, {\em in vitro} and {\em in vivo} studies showed that ESE--2 regulates the transcription of Snai2 \cite{chakrabarti2012elf5}.     
\end{description}

According to our model reduction methodology, literature search, careful manual curation, and network--based enrichment analysis; we propose that the derived core GRN module (see Figure 2) includes a molecular set which is both {\em necessary and sufficient} to specify the identity of the aforementioned cellular phenotypes and to represent the main intracellular regulatory events driving spontaneous immortalization in a robust manner. We test our proposal by building and analyzing a mechanistic GRN dynamical model (see below).

\subsection*{Recovered Attractors of the Core GRN Module Correspond to Configurations that Characterize Expected Cellular Phenotypes}

Based on the experimental data concerning the expression patterns of the genes incorporated in the proposed core GRN model in Figure 2 we assembled a table with a Boolean format of the state configurations expected to be recovered with the proposed GRN dynamical model. We refer to this configurations as the {\em ``expected attractors''} -- these correspond to the empirically observed genetic configurations. Furthermore, we integrated and formalized the experimental data concerning the interactions among the GRN nodes using Boolean logical functions that will rule the Boolean GRN dynamics and comprise the architecture of the proposed GRN. The set of formulated rules underlying the regulatory events is shown in Text S1 -- each logical rule is presented both as a logical preposition and as a truth table. Using the set of nodes and their corresponding logical rules we completely define a mechanistic dynamical GRN model \cite{davila2015descriptive}. The exhaustive computer--based simulation analysis of this model (see Methods) recovered three fixed--point attractors. Interestingly, the recovered attractors showed perfect correspondence with the expected attractors representing cellular phenotypes (see Table 2). The three recovered attractors correspond to the expected epithelial, senescent, and mesenchymal stem--like phenotypes :

\paragraph{The normal epithelial cell phenotype} is represented by the attractor with ESE--2, E2F, Cyclin and NF--$\kappa$B activity. ESE--2 is an epithelial--specific TF which regulates a large number of genes specific to epithelial cells \cite{siegel2013cancer,jemal2011global}. NF--$\kappa$B shows ubiquitous expression through the different types of human cells; however, it is also positively regulated by TFs of the ESE family ({\em i.e.} ESE--2) \cite{stratton2009cancer}. Moreover,  under inflammatory conditions the activity of  NF--$\kappa$B is enhanced \cite{hudson2010international,weinstein2013cancer}. On the other hand, E2F and Cyclin represent core regulators of cell--cycle entrance, and thus specify proliferative capability \cite{yaffe2013scientific,creixell2012navigating}.

\paragraph{The senescent cell phenotype} is represented by the attractor with ESE--2, Rb, p16, p53, and NF--$\kappa$B activity. Its biological counterpart would be an epithelial cell induced to replicative senescence, given (1) that it is expected to repress E2F \cite{siegel2013cancer}; and (2) that Rb, p16, p53, and NF--$\kappa$B are the molecular biomarkers of cellular senescence \cite{depinho2004cancer}.

\paragraph{Messenchymal Stem-like phenotype} In the model proposed here, the attractor whose configuration shows Snai2, Cyclin, NF--$\kappa$B, and Telomerase activity -- and inactivity of ESE--2, p16, Rb, p53, and E2F -- would correspond to a mesenchymal stem--like phenotype with tumorigenic potential (see discussion below). \\ 

The correspondence between the recovered attractors and the expected cellular phenotypes strongly suggests that the proposed nine--node core GRN indeed constitutes a regulatory module that is robust to initial conditions and that comprises a set of necessary and sufficient components and interactions to restrict the system to converge to the cellular phenotypes observed during spontaneous immortalization. 

\subsection*{Validation of the Uncovered Core Regulatory Module: Loss and gain--of--function Mutant and Robustness Analyses}

In order to validate the Boolean GRN dynamics we tested if the same GRN module is able to recover observed attractors in loss and gain of function mutants. We simulated such mutants analogous to experimental observations reported in the literature. Specifically, we simulated loss-- and gain--of--function mutations of ESE--2, Snai2, and p16 that have been reported in the literature. When simulating ESE--2 gain of function (by setting the expression state for this node permanently to ``1" in the simulations), the GRN model recovers three attractors corresponding to three different phenotypes which have been experimentally described and are associated with ESE--2 over--expression: an epithelial senescent cell \cite{fujikawa2007ese}, a normal epithelial cell \cite{chakrabarti2012elf5}, and a metastable state with proliferative phenotype \cite{lee2006epithelial}. In the case of ESE--2 loss--of--function (simulated by setting the expression state of this node to ``0" permanently), the model recovers an attractor corresponding to a mesenchymal phenotype, which is also consistent with observations \cite{chakrabarti2012elf5}. For Snai2, gain--of--function simulation recovers one attractor corresponding to mesenchymal stem--like phenotype, which is consistent with observations from ectopic over--expression experiments of mesenchymal TFs \cite{battula2010epithelial, cano2000transcription, sun2014slug}. Snai2 loss--of--function simulation, on the other hand, recovered two attractors corresponding to normal and senescent epithelial phenotypes, which is also consistent with observations \cite{chakrabarti2012elf5, liu2008zeb1}. Finally, gain--of--function simulation of p16 recovered two attractors; one associated with a mesenchymal stem--like but incompletely senescent (due to the lack of p53) phenotype; the other corresponding to an epithelial senescent phenotype. The first prediction is consistent with the status of immortal and apoptosis--resistant shown by mesenchymal stem--like cells, as well as with the capability of mesenchymal TFs to abrogate senescence \cite{weinberg2008twisted}. The second attractor is consistent with the potential for replicative senescence of epithelial cells. p16 loss--of--function simulation recovers two attractors corresponding to an epithelial cell and a mesenchymal stem--like cell. This prediction is consistent with the observed biological conditions for both phenotypes, where p16 is commonly repressed by polycomb proteins \cite{kim2006regulation}. The recovered attractors in mutant conditions are shown in Figure S2 in supplementary file. \\

It is important to note that, given that the uncovered regulatory module uncovered here is the result of a model reduction methodology where we permissively chose to represent multiple molecular species by the activity of some of the nodes, a direct interpretation of mutant simulations is not straightforward. Consequently, care should be taken when interpreting the results of the simulations or making predictions of mutant phenotypes yet to be experimentally tested and further explored in the context of the larger GRN in Figure 1, which is the focus of an ongoing study. With this in mind, instead of simulating additional mutant conditions, we further validated the dynamical GRN model by testing its robustness to perturbations of the logical rules. Specifically, we tested the robustness of the predicted attractors by generating a large set of perturbed networks ({\em e.g}, 10,000), calculating their respective attractors, and then counting the occurrences of the original attractors within the perturbed set. We generated each perturbed network by choosing a function of the network at random and flipping a single bit in this function \cite{mussel2010boolnet}. We performed four complementary {\em in silico} based experiments following this general robustness analysis. First, we estimated the fraction of occurrences of the three original attractors ({\em i.e.}, their robustness). Then, we repeated the experiment three times, but each time estimating the robustness of each individual attractor. For these four experiments we estimated a robustness ({\em i.e.}, fraction of times) of 0.7439, 0.905, 0.923, and 0.902, respectively. Hence, out of 10,000 random networks generated by {\em in silico} perturbations to the logical rules, a major fraction recovered the original attractors; as it is expected for a developmental (core) regulatory module that is robust both to transient (initial) and genetic perturbations \cite{espinosa2004gene}. This result supports the view that the core GRN uncovered here is indeed a regulatory network module driving developmental dynamics. It also constitutes a mechanistic explanation (for definitions, see \cite{davila2015descriptive}) to the generic cell phenotypes observed during spontaneous immortalization {\em in vitro} and which correlate with the cellular description of carcinoma progression {\em in vivo} (see below).

\subsection*{Attractor Time--Ordered Transitions: Epigenetic Landscape of the Uncovered GRN Core Module}

During the tumorigenic transformation of epithelial cells in culture, a generic time--ordered series of cell state transitions is observed and robustly induced by inflammation \cite{mani2008epithelial, battula2010epithelial}. Normal epithelial cell become senescent cells, which afterwards overcome this latter state acquiring a final mesenchymal stem--like phenotype. Interestingly, during the progressive pathological description of epithelial carcinomas {\em in vivo} the temporal pattern with which each of these different cell phenotypes enriches the tissue seems to be tightly ordered and is also generic to all types of such cancers irrespective of the tissue where they first appear. In order to test if the uncovered GRN core module not only underlies and restricts the types of cell phenotypes (attractors) but also their time--ordered transitions, following \cite{davila2015modeling} we explored its associated Epigenetic Landscape (EL) by implementing a discrete stochastic model as an extension to the Boolean network model \cite{alvarez2010abc} (see Methods). By means of computer--based simulations we performed two analyses in order to uncover functional and structural constraints in attractor transitions. (1) We explored the temporal sequence of attractor attainment, and (2) we calculated the consistent global ordering of all the given attractors. Specifically, following \cite{alvarez2008floral}, we found that the most probable temporal order of attractor attainment for a cell (population) initially on epithelial state is:
\[
\mbox{Epithelial}\rightarrow\mbox{Senescent}\rightarrow\mbox{Mesenchymal stem--like},
\]
see Fig 4a. On the other hand, following \cite{zhou2014relative} we defined a consistent global ordering of the uncovered attractors based on their relative stability (see Methods). Relative stability calculations are based on the mean first passage time (MFPT) between pairs of attractors. These, in turn, epitomize barrier heights in the EL by approximating a measure for the ease of specific transitions. Similar to the previous analysis, the uncovered global ordering of attractors is Epithelial $\rightarrow$ Senescent $\rightarrow$ Mesenchymal stem--like (Fig 4b). This corresponds to the only order in which the system can visit the three attractors following a positive net transition rate. These results indicate that, when considering only intracellular regulatory constraints alone, the uncovered GRN core module structures the epigenetic landscape in a way that a specific flow across the landscape is preferentially and robustly followed. We anticipate that observed transition rates in vivo are likely to depend on tissue--level processes and/or additional GRN components underlying epithelial cell sub--differentiation, that have not been considered here. These latter restrictions will be modeled in future contributions building up on the framework that has been put forward here.  

\section*{Discussion}

Multicellularity by definition implies a one--to--many genotype--phenotype map. The genome of a multicellular individual possesses the intrinsic potentiality to implement a developmental process by which all its different cell--types and tissue structures are ultimately established. In the last decades, a quite coherent theory to explain the development of multicellular organisms as the result of the orchestrating role of GRNs has been developed \cite{mendoza1998gene,huang2009complex, alvarez2010abc}.  The main conclusion is that observable cell states emerge from the self--consistent multistable regulatory logic dictated by genome structure and obeyed by (mainly) transcription factors (TFs) resulting in stable, steady--states of gene expression. Cancer development and progression is also a phenomenon intrinsic to multicellular organisms. Furthermore, similar to normal development, cancer is robustly established as evidenced by its directionality and phenotypic convergence \cite{huang2011intrinsic}. Is cancer somehow orchestrated by GRN dynamics as well? Several hypothesis have been presented in this direction such as the cancer attractor theory \cite{huang2009cancer, huang2011intrinsic}, and the endogenous molecular cellular network hypothesis \cite{wang2014quantitative, zhu2015endogenous}. In this contribution we also follow the viewpoint of an intrinsic regulatory network, but we focus on a specific developmental process at the cellular level: the robust cell state transitions observed during the tumorigenic transformation of human epithelial cells in culture induced by inflammation and resulting from surpassing a senescent state through EMT -- {\em i.e.,} tumorigenic transformation due to spontaneous immortalization. We propose that a mechanistic understanding of this process is an important first necessary step to unravel key cellular processes which might be occurring {\em in vivo}, where its rate of occurrence is likely to be regulated by tissue--level and systemic conditions directly linked with lifestyle choices, as well as additional regulatory interactions underlying epithelial cell sub--differentiation.

\subsection*{A Generic Molecular Regulatory Network} 

The predominant strategy in the molecular study of cancer and cellular tumorigenic transformation has been to focus on pathways and associated mutations. Aware that signaling pathways are actually embedded in complex regulatory networks here we assembled from curated literature a GRN comprising the main molecular regulators involved in key cellular processes ubiquitous to carcinogenesis following a bottom--up approach (see results). Subsequently, we followed a mechanistic approach to address the question of whether we assembled a set of necessary and sufficient molecular players and interactions to recover the cellular phenotypes and processes documented during the spontaneous immortalization of human epithelial cells in culture: we proposed, analyzed and validated an experimentally grounded core GRN dynamical model. \\

Small developmental regulatory modules have been shown to successfully include the necessary and sufficient set of components and interactions for explaining, as manifestations of intrinsic structural and functional constraints imposed by these GRNs, the dynamics of complex processes such as stem cell differentiation \cite{li2013quantifying}, cell--fate decision \cite{zhou2011predicting} and similar cellular processes during plant morphogenesis \cite{mendoza1998gene,espinosa2004gene, alvarez2008floral, azpeitia2010single}. We hypothesized that a similar core developmental module can be formulated in an attempt to explain the cell--fates observed  during spontaneous immortalization of human epithelial cells {\em in vitro} resulting in a potentially tumorigenic state. In order to show this, we first reduced the proposed larger network into a regulatory core module, by eliminating transitory pathways within the network and by including compounded nodes while maintaining the core network structure and without affecting the dynamical output during each reduction step (for details, see Methods). We obtained a small set of main molecular players (Fig 2). We extracted from available literature the expression profiles of the generally observable cell states of interest in terms of this minimal set of molecules (see Table 2). Given our main hypothesis, we tested if the reduced molecular set and their regulatory logic formalized as a Boolean GRN model were able to recover the biologically observable expression profiles as stationary and stable network configurations ({\em i.e.}, attractors). Interestingly, we found that the core GRN model only converges to the observed gene expression profiles in wild--type (see Table 2) and some mutant backgrounds (see results). This result strongly suggest that we have successfully included the key regulators and interactions at play during the establishment of cell states observed during the tumorigenic transformation of human epithelial cells resulting from spontaneous immortalization. \\

It is noteworthy that our model does not include any hypothetical interaction or component, a common practice in GRN modeling \cite{espinosa2004gene,azpeitia2010single, zhou2011predicting}. Our GRN model exclusively integrates available published experimental data; indeed, it was a surprising result that the observed dynamical behavior emerged naturally under such conditions. This suggests that despite incomplete information, there is enough molecular data to uncover important restrictions underlying cell behavior during transitions relevant to epithelial carcinogenesis. Consequently, we consider that the networks reported herein (both the large and the core GRNs) may serve as {\em bona fide} base models useful to integrate novel discoveries, as well as components underlying epithelial cellular sub--differentiation, while following a bottom--up approach in cancer network systems biology.  

\subsection*{Attractor Time--Ordered Transitions} 

Discrete GRN models can be used to integrate regulatory mechanisms that not only recapitulate the observed gene expression patterns, but that also reproduce the observed developmental time--ordering of cell phenotypes. This can be done by considering stochasticity in the model in order to explore \cite{alvarez2010abc, azpeitia2014gene, davila2015modeling} and/or characterize \cite{zhou2014relative} the associated EL. Importantly, by exploring noise--induced transitions we do not assume that noise alone is the driving force of the transitions, instead, we exploit noise as a tool to explore the GRN--based version of Waddington's EL and to indirectly characterize its structure. Specifically, by calculating the relative stability of the attractors (see Methods) we approximate the in--between attractor barrier heights in the landscape. Furthermore, measures of relative stability can also be exploited to calculate net transition rates measuring the ease of specific inter--attractor transitions and to uncover the predominant developmental route across the epigenetic landscape \cite{Zhou2014Discrete}: ordered transitions sharing positive net transition rates will be preferentially followed. Our results show that such a developmental route follows the time--order of cellular phenotypic states epithelial$\rightarrow$senescent$\rightarrow$mesenchymal stem--like (potentially tumorigenic). In other words, the constraints imposed by the GRN structure the associated EL in such a way that an epithelial cell in culture as a ``ball" would naturally roll following such a path, in agreement with the observed spontaneous immortalization process. \\

Even in the case of the simple model presented here, it is interesting that of the many possible cell states and developmental routes, the core GRN network is canalized to the few steady--states and the developmental time--ordering consistent with the molecular characterization of cell phenotypes observed during spontaneous immortalization and correlating with carcinoma progression {\em in vivo} (see below). This suggests that specific progressive alterations or particular ``abnormal'' signaling mechanisms are not necessarily required for a cell to reach a potentially tumorigenic state. Additionally, robustness analysis performed on the same network showed that the recovered attractors are also robust to permanent alterations of the regulatory logic.

\subsection*{From Abstract Network Attractors and Dynamics to Biological Insight}

We are aware of the high degree of simplification involved in the model proposed herein. Accordingly, we do not attempt to present it as a source of accurate predictions for either the occurrence or the future behavior of a phenomena as complex as carcinogenesis. Instead, we formulate the model in an attempt to provide some intuition into otherwise highly complicated processes, and to illuminate increasing body of confounding descriptions. Simple mechanistic models like the one presented here sacrifice detail and accuracy in exchange for understanding \cite{lander2010edges, davila2015descriptive}. What biological insights can be gained by the uncovered GRN dynamical model? Our simple GRN model strongly suggests that the generic series of cell state transitions widely observed and robustly induced by inflammation in cell culture from normal epithelial to immortalized senescent cells, and from this latter state to a final mesenchymal stem--like phenotype in the process defined as spontaneous immortalization naturally result from the self--organized behavior emerging from an underlying GRN novel architecture and topology. \\

Importantly, cells that emerge from spontaneous immortalization induced by cytokines display mesenchymal stem like phenotype and tumorigenic behavior -- {\em i.e.}, repress proteins p16 and p53, surpass senescence, and re--express telomerase \cite{battula2010epithelial}. Phenotypically, these cells are difficult to distinguish from the so--called cancer stem cells, tumor initiating cells or embryonic stem cells \cite{morel2008generation,neph2012circuitry}; are resistant to apoptosis; and have the ability to migrate and generate metastasis and form secondary tumors -- all lethal traits characterizing cancer cells \cite{mani2008epithelial}. We, thus, speculate that tissue--level conditions associated with a bad prognosis, such as a pro--inflammatory milieu, may increase the rate of occurrence of these same transitions {\em in vivo} promoting as a result the development and progression of epithelial cancer. We substantiate this view by noting several independent empirical observations. (1) Histological diagnosis of carcinoma are generally preceded by a lesion called hyperplasia; senescent cells are abundant in hyperplasias and scarce in carcinomas \cite{chen2005crucial}. (2) During chronological aging senescent cells increase in number within both normal tissues and hyperplasias. (3) Senescence is associated with the promotion of carcinogenesis by contributing with the loss of tissue architecture and promoting an inflammatory milieu \cite{campisi2011cellular}. (3) Overcoming the senescent barrier is fundamental in tumor progression \cite{narita2005senescence, yildiz2013genome}. (4) The EMT process constitutes a well--characterized mean to overcome senescence under an inflammatory environment(\cite{smit2010epithelial}). \\

We must point out, however, that transition rates during spontaneous immortalization, if occurring {\em in vivo}, may be regulated by tissue--level, self--organizational processes not considered in our cellular level model. For example, the likelihood of spontaneous immortalization {\em in vivo} may be increased by extracellular perturbations that inevitably occur during aging; mainly, by inflammation and tissue remodeling resulting from an increased population of senescent cells. The cellular level network models reported here are, nevertheless, a valuable building block for more detailed multi--level models integrating further sources of tissue--level constraints such as cell cycle progression, cell--cell interactions, differential proliferation rates, and mechanical forces. \\

Summarizing, in this contribution we propose an experimentally grounded GRN model for spontaneous immortalization. We report one large GRN model (41 nodes) and one core GRN developmental module (9 nodes), both useful and necessary for further integration of signaling and mechanical processes in multi--level, more detailed modeling efforts. We explore by analyzing the dynamical behavior of the latter if the uncovered GRN topology and architecture underlies the gene expression configurations that characterize normal epithelial, senescent, and mesenchymal stem--like cell--fates well documented during tumorigenic transformation {\em in vitro} and which correlate with those observed in the progressive pathological description of epithelial carcinogenesis {\em in vivo}. Overall, our results suggest that tumorigenic transformation {\em in vitro} due to spontaneous immortalization can be understood and modeled at a cellular level generically as a developmental system undergoing cell--state transitions resulting from the structural and functional constraints imposed, in part, by the interactions included in the proposed GRN. They also suggest that similar transitions may be occurring {\em in vivo} and might be relevant for carcinoma development and progression. This view is consistent with the robustness, generic patterns, and directionality observed during the development of human cancers derived from epithelial tissues. Particularly, based on our results, we hypothesize that replicative senescence and chronic inflammation are likely to increase the occurrence of spontaneous immortalization {\em in vivo} promoting the development of epithelial carcinogenesis. Testing such hypothesis awaits the development of multi--level models taking the ones presented here as building blocks, which is the subject of ongoing investigation.

\section*{Materials and Methods}

\subsection*{Literature Search}

A total of 159 references, considering both references in extended view material (see Text S1) and main text, were carefully and manually reviewed in order to first define a minimal set of cellular phenotypes and processes (for definitions, see Text S1) which enable a generic representation of  epithelial carcinogenesis on the basis of cell state transition events. Subsequently, a set of  associated, experimentally described molecular regulators was extracted from the literature, including their regulatory interactions.

\subsection*{Network Assembly}

The network (see Fig. 1) was assembled manually by adding nodes (genes/proteins) and edges (activating or inhibitory interactions) describing direct mechanisms reported in the available literature to have an influence on both the specification of the cellular phenotypes and the development of the cellular process defined in (Text S1). The initial network was created based on experimentally grounded knowledge from 159 references (including reviews and research papers) and consists of 41 nodes and 97 edges. The literature included data known before 2014. Support for each of the proposed interactions is listed in Text S1.

\subsection*{Network--based Gene Set Enrichment Analysis}

The bioinformatics tools EnrichNet \cite{glaab2012enrichnet} and TopoGSA \cite{glaab2010topogsa} were used to perform network--based gene set enrichment analysis and topology--based gene set analysis, respectively. Briefly, EnrichNet maps the input gene set into a molecular interaction network and calculates distances between the genes and pathways/processes in a reference database. TopoGSA also maps the input gene set into a network, and then it computes its topological statistics and compares it against the topology of pathways/processes in a reference database. Here a connected human interactome graph extracted from the STRING database and the KEGG and GO Biological Process databases were used as reference molecular interaction network and databases. Both analyses were performed using the Cytoscape plugin Jepettp \cite{winterhalter2014jepetto}.

\subsection*{Network Reduction}

In order to extract a representative core regulatory model from the initial network and to obtain a more computationally tractable one, which reasonably unfolds the regulatory pathways, a reduction methodology was followed based on certain simplifying assumptions -- supported by previous results in molecular biology studies -- and on mathematical results from dynamical systems and graph theory. Here we briefly describe the main steps. The step--by--step reduction process is included in Text S1.

\paragraph{Simplifying assumptions:}
\begin{itemize}
	\item ESE-2 groups activities of  ESE-1, ESE-3, EGF, Her-2/neu.
	\item Snai2 groups activities of Snail, Twist (Twist, in turn, groups activities of Twist1 and Twist2), Zeb and FOXC2.
	\item p16 groups p14 and NF--$\kappa$B node groups the inflammatory response activated by growth factors, mitogens and cytokines.
\end{itemize}

\paragraph{Reduction process:}

(1) Simple mediator nodes ({\em i.e.}, those nodes with in--degree and out--degree of one) were removed iteratively. (2) Nodes with in--degree of one and out--degree greater than one were removed iteratively. These steps (1 and 2) does not alter the attractors of the Boolean network under the asynchronous update, as mathematically proved in \cite{saadatpour2013reduction}. (3) Redundant interactions of selected nodes (based on biological arguments) resulting in self--regulation were included in single nodes/interactions (for details, see Text S1). (4) Selected nodes (based on biological knowledge again) with in--degree greater than one and out--degree of one were removed. The final steps (3 and 4) are supported by the mathematical analysis made in \cite{naldi2011dynamically} in which the authors prove that the methodology preserves relevant topological and dynamical properties.

It is noteworthy that fixed point attractors are time--independent, so they are the same in both synchronous and asynchronous update methods. Complex attractors (in which the system oscillates among a set of states), on the other hand, depend on the update method. Consistently, the update method used in the model is irrelevant for the obtained results. This last assertion is valid because the model shows only fixed point attractors, which means, under the mathematically proved reduction methods applied, that {\em the large network describes a qualitative long time behavior conserved in the reduced one}. Besides, the methodology applied in order to obtain the reduced network enables the analysis of a resulting regulatory graph which is biologically meaningful and dynamically consistent with the network constructed with available molecular biology experimental data.

The final reduced network is shown in Figure 2. We refer to this network and its corresponding logical rules as the core regulatory module.

\subsection*{Dynamical Gene Regulatory Network Model}

A Boolean network models a dynamical system assuming both discrete time and discrete state variables. This is expressed formally with the mapping:
\begin{equation}
x_i(t+1) = F_i(x_1(t),x_2(t),...,x_k(t)),
\end{equation}
where the set of functions $F_i$ are logical prepositions (or truth tables) expressing the relationship between the genes that share regulatory interactions with the gene $i$, and where the state variables $x_i(t)$ can take the discrete values $1$ or $0$ indicating whether the gene $i$ is expressed or
not at a certain time $t$, respectively. An experimentally grounded Boolean GRN model is then completely specified by the set of genes proposed to be involved in the process of interest and the associated set of logical functions derived from experimental data \cite{azpeitia2014gene}. The set of logical functions for the core regulatory module used in this study is included in Text S1 -- both as logical prepositions and truth tables. The dynamical analysis of the Boolean network model was conducted using the package {\em BoolNet} \cite{mussel2010boolnet} within the {\em R} statistical programming environment (www.R-project.org).

\subsection*{Epigenetic Landscape Exploration}

\subsubsection*{Including Stochasticity}

In order to extend the Boolean Network into a discrete stochastic model and then study the properties of its associated EL, the so--called stochasticity in nodes (SIN) model was implemented following \cite{alvarez2008floral, azpeitia2014gene, davila2015modeling}. In this model, a constant probability of error $\xi$ is introduced for the deterministic Boolean functions. In other words, at each time step, each gene ``disobeys'' its Boolean function with probability $\xi$. Formally:
\begin{equation}
\begin{aligned}
&  P_{x_i(t+1)}[F_i(\mathbf{x}_{reg_i}(t))] = 1- \xi, \\
& P_{x_i(t+1)}[1 - F_i(\mathbf{x}_{reg_i}(t))] = \xi.
\end{aligned}
\end{equation}
The probability that the value of the now random variable $x_i(t+1)$ is determined or not by its associated logical function $F_i(\mathbf{x}_{reg_i}(t))$ is $1- \xi$ or $\xi$, respectively.  

\subsubsection*{Attractor Transition Probability Estimation}

An attractor transition probability matrix $\Pi$ with components: 
\begin{equation}
\pi_{ij} = P(A_{t+1}=j|A_t=i),
\end{equation}
representing the probability that an attractor $j$ is reached from an attractor $i$, was estimated by numerical simulation following \cite{alvarez2008floral}. Specifically, for each network state $i$ in the state space ($2^n$) a stochastic one--step transition was simulated a large number of times ($\approx 10,000$). The probability of transition from an attractor $i$ to an attractor $j$ was then estimated as the frequency of times the states belonging to the basin of the attractor $i$ were stochastically mapped into a state within the basin of the attractor $j$.    

Following the discrete time Markov chains (DTMCs) \cite{allen2010introduction} theoretical framework, the estimated transition probability matrix was integrated into a dynamic equation for the probability distribution:
\begin{equation}
P_A(t+1) = \Pi P_A(t), 
\end{equation}
where $P_A(t)$ is the probability distribution over the attractors
at time $t$, and $\Pi$ is the transition probability matrix. This equation was iterated to simulate the temporal evolution of the probability distribution over the attractors starting from a specific initial probability distribution.

\subsubsection*{Attractor Relative Stability and Global Ordering Analyses}

In addition to the calculation of the most probable temporal cell--fate pattern (see \cite{alvarez2008floral}), a discrete stochastic GRN model enables the study of the ease for transitioning from one attractor to another \cite{Zhou2014Discrete}. Specifically, a transition barrier in the EL epitomizes the ease for transitioning from one attractor to another. The ease of transitions, in turn, offers a notion of relative stability. It has recently been proposed that the GRN has a consistent global ordering of all cell attractors and intermediate transient states which can be uncovered by measuring the relative stabilities of all the attractors of a Boolean GRN \cite{Zhou2014Discrete, zhou2014relative}. Here, the relative stabilities of the cell states were defined based on the mean first passage time (MFPT). Specifically, a relative stability matrix $M$ was calculated which reflects the transition barrier between any two states based on the MFPT. Here, in all cases, the MFPT was estimated numerically. Using the transition probabilities among attractors, a large number sample paths of a finite Markov chain were simulated. The MFPT from attractor $i$ to attractor $j$ corresponds to the averaged value of the number of steps taken to visit attractor $j$ for the fist time, given that the entire probability mass was initially localized at the attractor $i$. The average is taken over the realizations. Following \cite{Zhou2014Discrete}, based on the  MFPT values a  net transition rate between attractor $i$ and $j$ can be defined as follows:
\begin{equation}
d_{i,j} = \frac{1}{MFPT_{i,j}} - \frac{1}{MFPT_{j,i}}
\end{equation}
This quantity effectively measures the ease of transition as a net probability flow. For all the calculation involving stochasticity, the robustness of the results was assessed by taking three different values for the probability of error  $\left(0.01, 0.05, 0.1\right)$. Stability of the results was assessed by manually changing the number of simulated samples until results become stable. 

The consistent global ordering of all attractors uncovered with the core GRN was defined based on the formula proposed in \cite{zhou2014relative}. Briefly, the consistent global ordering of the attractors is given by the attractor permutation in which al transitory net transition rates from an initial attractor to a final attractor are positive. This is schematically represented in Figure 4b.  Calculated transition probability, MFPT, and net transition rate matrices are included in Text S2. R source code implementing all the calculations and analyses is available upon request.

\section*{Authors' contributions}

ERAB and JMG coordinated the study and with the other authors established the overall logic and core questions to be addressed. All the authors conceived and planned the modeling approaches. FML recovered the information from the literature to establish the model and provided expert knowledge in cancer biology. JDV established many of the specific analyses to be done, and programmed and ran all the modeling and analyses. FML and CEO formalized experimental data into regulatory logic. ERAB, JMG, JDV and FML participated in the interpretation of the results and analyses. JDV wrote most of the paper with help from ERAB and JMG and input from FML. All authors proofread the final version of the ms submitted.

\section*{Acknowledgments}
This work was supported by grants CONACYT 240180, 180380, 167705, 152649 and UNAM-DGAPA-PAPIIT: IN203113, IN 203214, IN203814, UC Mexus ECO-IE415. J.D.V acknowledges the support of CONACYT and the Centre for Genomic Regulation (CRG), Barcelona, Spain; while spending a research visit in the lab of Stephan Ossowski. This article constitutes a partial fulfillment of the graduate program Doctorado en Ciencias Biom\'edicas of the Universidad Nacional Aut\'onoma de M\'exico, UNAM in which J.D.V. developed this project. J.D.V receives a PhD scholarship from CONACYT. The authors acknowledge logistical and administrative help of Diana Romo.



\bibliography{Cancer_MS_150417.bib}

\section*{Tables}

\begin{table}[h]
	\begin{tabular}{|c|c|c|c|} \hline
		{\bf KEGG -- Pathway or Process} & {\bf XD--score} & {\bf q-value} & {\bf Overlap/Size} \\ \hline
		{\em Bladder cancer} & 1.1447 & 0 &   12/38 \\
		Chronic myeloid leukemia &	0.86866	& 0	 & 17/69 \\
		p53 signaling pathway &	0.78477	 & 0	&  14/62 \\
		{\em Pancreatic cancer} &	0.68155	& 0	&  14/70 \\
		Glioma &	0.68155 &	0 &	  12/60 \\
		{\em Non--small cell lung cancer} &	0.66586	& 0	&  10/51 \\ 
		Melanoma &	0.65574 &	0	&  12/62 \\
		{\em Small cell lung cancer} &	0.56447 &	0 &	  14/82 \\
		{\em Prostate cancer} &	0.54821 &	0	&  14/84 \\
		Cell cycle &	0.54821 &	0 &	 20/120 \\
		Cytosolic DNA--sensing pathway &	0.48155 &	0.00001 &   6/40 \\
		Thyroid cancer &	0.36155 &	0.00784 &   3/25 \\
		NOD-like receptor signaling pathway & 0.35612 &	0.00001 &   7/59 \\ \hline
		{\bf GO Biological Process} &	{\bf XD-score} &	{\bf q--value} &	{\bf Overlap/Size} \\ \hline
		{\em replicative senescence}	& 3.13328 & 0 	&   8/10 \\
		cellular senescence	 & 0.73328 &	0.02244 &	  2/10 \\
		cell aging &	0.43328	 & 0.00608	 &  3/24 \\
		activation of NF--$\kappa$B--inducing kinase activity &	0.43328 &	0.04656	   & 2/16 \\
		determination of adult lifespan	 & 0.33328	& 0.40382	 &  1/10 \\
		{\em epithelial cell differentiation} &	0.32721 &	0.13188	&   2/33 \\ 
		{\em positive regulation of NF--$\kappa$B transcription factor activity} &	0.30109	& 0	 &  8/87 \\ \hline
	\end{tabular}
	\begin{flushleft} 
	\end{flushleft}
	\caption{
		\bf{Significant pathways and processes according to network--based gene set enrichment analysis}}
	\label{tab:label}
\end{table}

\begin{table}[h]
	{\small 
		\begin{tabular}{|c|c|c|c|} \hline
			{\bf Cellular Phenotype} & {\bf Recovered Attractor (Active)} & {\bf {\em``Expected Attractors''}} & {\bf References} \\ \hline
			Epithelial & {\color{red} Ese--2}, NF-$\kappa$B, {\color{blue} E2F, Cyclin} & {\color{red} Ese--2}, NF--$\kappa$B, {\color{blue} Cell Cycle(+)}& \cite{chakrabarti2012elf5} \\
			Senescent & {\color{red} p16, p53}, Ese--2, NF--$\kappa$B, {\color{blue} Rb} & {\color{red} p16, p53}, NF--$\kappa$B, {\color{blue} Cell Cycle(-)} & \cite{li2011regulatory, yamakoshi2009real, zeisberg2009biomarkers} \\
			Mesenchymal stem-like & {\color{red} Snai2, Telomerase}, NF--$\kappa$B, {\color{blue} Cyclin} & {\color{red} Snai2, Telomerase}, NF--$\kappa$B, {\color{blue} Cell Cycle(+)}  & \cite{zeisberg2009biomarkers, chakrabarti2012elf5} \\ \hline
		\end{tabular}
	}
	\begin{flushleft}
		\caption{
			\bf{Predicted and Observed Attractors}}
	\end{flushleft}
	\label{tab:label}
\end{table}

\section*{Figure Legends}

\noindent {\bf Figure 1. Gene regulatory network for epithelial carcinogenesis}. Nodes represent genes, and arrows (bars) represent experimentally characterized activation (arrow-heads) or repression (flat-heads) interactions. Genes corresponding to TFs are represented by squares and the rest by circles. (a) Colors indicate association with specific phenotypes and processes: epithelial (green), mesenchymal (orange), inflammation (red), senescence and DNA damage (blue), cell--cycle (purple), and polycomb complex (yellow). (b) Core gene regulatory module in the context of the global network. Colored nodes represent the final set of molecules obtained after the network reduction methodology was applied (see Methods) and which were included in the core GRN model.

\vspace{1.0cm}

\noindent {\bf Figure 2. Core gene regulatory network module for epithelial carcinogenesis} Nodes represent either single or subsets of genes (see Results); arrows-heads represent activations and flat--heads repression interactions. Five of the nodes are involved in the specification of the cellular phenotypes: Epithelial (Ese--2), Senescent (p16, p53), and Mesenchymal stem--like (Snai2, TELasa). Three nodes are tightly associated with cell--cycle regulation (Rb, E2F, Cyclin),  while node NF--$\kappa$B represents cellular inflammation.

\vspace{1.0cm}

\noindent {\bf Figure 3.  The core gene regulatory module in the context of the} {\em Hallmarks of Cancer} approach. The antagonistic activity state ESE--2 (-) and Snai2 (+) enable cells to {\em sustain proliferative signals} and {\em evade growth suppressors} by undergoing a dedifferentiation process. The state p16(-), Rb(-), p53(-),  and TELasa (+) enable cell to {\em acquire replicative immortality}, {\em resist cell death}, as well as present {\em genome instability} and a {\em mutation--prone} phenotype by  surpassing cellular senescence. High levels of cytokines and NF--$\kappa$B(+) expose cells to {\em tumor promoting inflammation}. The constitutive activity of Snai2(+) epitomizes the intrinsic phenotypic features of the cells emerging from the process of inflammation--induced EMT: {\em activating invasion}, {\em avoiding immune destruction}, and {\em deregulating cellular energetics}.

\vspace{1.0cm}

\noindent {\bf Figure 4.   Temporal sequence and global order of cell--fate attainment pattern under the stochastic Boolean GRN model during epithelial carcinogenesis.}  (a) Maximum probability $p$ of attaining each attractor, as a function of time (in iteration steps). Vertical lines mark the time when maximal probability of each attractor occurs. The most probable sequence of cell attainment is: epithelial(E) $\rightarrow$ senescent(S) $\rightarrow$ mesenchymal(cancer--like)(M). The value of the error probability used in this case was $\xi=0.05$. The same patterns were obtained with the 3 different error probabilities tested (data not shown). (b) Schematic representation of the possible transitions between pairs of attractors. Arrows indicate the directionality of the transitions. Above each arrow a sign ($+$) or ($-$) indicates whether the calculated net transition rate between the corresponding attractors is positive or negative. Red arrows represent the globally consistent ordering for the 3 attractors: the order of the attractors in which all individual transition has a positive net rate, resulting in a global probability flow across the EL.


\bibliographystyle{vancouver}

\bibliography{Cancer_MS_150417}

\begin{thebibliography}{10}

\bibitem{anand2008cancer}
Anand P, Kunnumakara AB, Sundaram C, Harikumar KB, Tharakan ST, Lai OS, et~al.
\newblock Cancer is a preventable disease that requires major lifestyle
  changes.
\newblock Pharmaceutical research. 2008;25(9):2097--2116.

\bibitem{huang2011intrinsic}
Huang S.
\newblock On the intrinsic inevitability of cancer: from foetal to fatal
  attraction.
\newblock In: Seminars in cancer biology. vol.~21. Elsevier; 2011. p. 183--199.

\bibitem{mani2008epithelial}
Mani SA, Guo W, Liao MJ, Eaton EN, Ayyanan A, Zhou AY, et~al.
\newblock The epithelial-mesenchymal transition generates cells with properties
  of stem cells.
\newblock Cell. 2008;133(4):704--715.

\bibitem{huang2009non}
Huang S.
\newblock Non-genetic heterogeneity of cells in development: more than just
  noise.
\newblock Development. 2009;136(23):3853--3862.

\bibitem{ben2008embryonic}
Ben-Porath I, Thomson MW, Carey VJ, Ge R, Bell GW, Regev A, et~al.
\newblock An embryonic stem cell--like gene expression signature in poorly
  differentiated aggressive human tumors.
\newblock Nature genetics. 2008;40(5):499--507.

\bibitem{kelloff2007assessing}
Kelloff GJ, Sigman CC.
\newblock Assessing intraepithelial neoplasia and drug safety in
  cancer-preventive drug development.
\newblock Nature Reviews Cancer. 2007;7(7):508--518.

\bibitem{virchow1860cellular}
Virchow RLK.
\newblock Cellular pathology.
\newblock John Churchill; 1860.

\bibitem{huang2009cancer}
Huang S, Ernberg I, Kauffman S.
\newblock Cancer attractors: a systems view of tumors from a gene network
  dynamics and developmental perspective.
\newblock In: Seminars in cell \& developmental biology. vol.~20. Elsevier;
  2009. p. 869--876.

\bibitem{mendoza1998gene}
Mendoza L, Alvarez-Buylla ER.
\newblock Dynamics of the genetic regulatory network for Arabidopsis thaliana
  flower morphogenesis.
\newblock J Theor Biol. 1998;193(2):307--319.

\bibitem{espinosa2004gene}
Espinosa-Soto C, Padilla-Longoria P, Alvarez-Buylla ER.
\newblock A gene regulatory network model for cell-fate determination during
  Arabidopsis thaliana flower development that is robust and recovers
  experimental gene expression profiles.
\newblock The Plant Cell Online. 2004;16(11):2923--2939.

\bibitem{huang2009complex}
Huang S, Kauffman S.
\newblock Complex gene regulatory networks-from structure to biological
  observables: cell fate determination.
\newblock Encyclopedia of Complexity and Systems Science Meyers RA, editors
  Springer. 2009;p. 1180--1293.

\bibitem{alvarez2010abc}
Alvarez-Buylla ER, Azpeitia E, Barrio R, Ben{\'\i}tez M, Padilla-Longoria P.
\newblock From ABC genes to regulatory networks, epigenetic landscapes and
  flower morphogenesis: making biological sense of theoretical approaches.
\newblock Seminars in cell \& developmental biology. 2010;21(1):108--117.

\bibitem{huang2009reprogramming}
Huang S.
\newblock Reprogramming cell fates: reconciling rarity with robustness.
\newblock Bioessays. 2009;31(5):546--560.

\bibitem{kaneko2011characterization}
Kaneko K.
\newblock Characterization of stem cells and cancer cells on the basis of gene
  expression profile stability, plasticity, and robustness.
\newblock Bioessays. 2011;33(6):403--413.

\bibitem{huang2012molecular}
Huang S.
\newblock The molecular and mathematical basis of Waddington's epigenetic
  landscape: A framework for post-Darwinian biology?
\newblock Bioessays. 2012;34(2):149--157.

\bibitem{davila2014bridging}
Davila-Velderrain J, Alvarez-Buylla ER.
\newblock Bridging the Genotype and the Phenotype: Towards An Epigenetic
  Landscape Approach to Evolutionary Systems Biology.
\newblock bioRxiv. 2014;.

\bibitem{xu2009tgf}
Xu J, Lamouille S, Derynck R.
\newblock TGF-$\beta$-induced epithelial to mesenchymal transition.
\newblock Cell research. 2009;19(2):156--172.

\bibitem{battula2010epithelial}
Battula VL, Evans KW, Hollier BG, Shi Y, Marini FC, Ayyanan A, et~al.
\newblock Epithelial-Mesenchymal Transition-Derived Cells Exhibit Multilineage
  Differentiation Potential Similar to Mesenchymal Stem Cells.
\newblock Stem Cells. 2010;28(8):1435--1445.

\bibitem{li2012epithelial}
Li CW, Xia W, Huo L, Lim SO, Wu Y, Hsu JL, et~al.
\newblock Epithelial--mesenchymal transition induced by TNF-$\alpha$ requires
  NF-$\kappa$B--mediated transcriptional upregulation of Twist1.
\newblock Cancer research. 2012;72(5):1290--1300.

\bibitem{morel2008generation}
Morel AP, Li{\`e}vre M, Thomas C, Hinkal G, Ansieau S, Puisieux A.
\newblock Generation of breast cancer stem cells through epithelial-mesenchymal
  transition.
\newblock PloS one. 2008;3(8):e2888.

\bibitem{neph2012circuitry}
Neph S, Stergachis AB, Reynolds A, Sandstrom R, Borenstein E,
  Stamatoyannopoulos JA.
\newblock Circuitry and dynamics of human transcription factor regulatory
  networks.
\newblock Cell. 2012;150(6):1274--1286.

\bibitem{azpeitia2010single}
Azpeitia E, Ben{\'\i}tez M, Vega I, Villarreal C, Alvarez-Buylla ER.
\newblock Single-cell and coupled GRN models of cell patterning in the
  Arabidopsis thaliana root stem cell niche.
\newblock BMC systems biology. 2010;4(1):134.

\bibitem{azpeitia2014gene}
Azpeitia E, Davila-Velderrain J, Villarreal C, Alvarez-Buylla ER.
\newblock Gene regulatory network models for floral organ determination.
\newblock In: Flower Development. Springer; 2014. p. 441--469.

\bibitem{alvarez2008floral}
{\'A}lvarez-Buylla ER, Chaos {\'A}, Aldana M, Ben{\'\i}tez M, Cortes-Poza Y,
  Espinosa-Soto C, et~al.
\newblock Floral morphogenesis: stochastic explorations of a gene network
  epigenetic landscape.
\newblock Plos one. 2008;3(11):e3626.

\bibitem{davila2015modeling}
Davila-Velderrain J, Mart{\'\i}nez-Garc{\'\i}a J, Alvarez-Buylla ER.
\newblock Modeling the Epigenetic Attractors Landscape: Towards a Post-Genomic
  Mechanistic Understanding of Development.
\newblock Name: Frontiers in Genetics. 2015;6:160.

\bibitem{zhou1998novel}
Zhou J, Ng A, Tymms MJ, Jermiin LS, Seth AK, Thomas RS, et~al.
\newblock A novel transcription factor, ELF5, belongs to the ELF subfamily of
  ETS genes and maps to human chromosome 11p13-15, a region subject to LOH and
  rearrangement in human carcinoma cell lines.
\newblock Oncogene. 1998;17(21):2719--2732.

\bibitem{ma2003gene}
Ma XJ, Salunga R, Tuggle JT, Gaudet J, Enright E, McQuary P, et~al.
\newblock Gene expression profiles of human breast cancer progression.
\newblock Proceedings of the National Academy of Sciences.
  2003;100(10):5974--5979.

\bibitem{escamilla2010genome}
Escamilla-Hernandez R, Chakrabarti R, Romano RA, Smalley K, Zhu Q, Lai W,
  et~al.
\newblock Genome-wide search identifies Ccnd2 as a direct transcriptional
  target of Elf5 in mouse mammary gland.
\newblock BMC molecular biology. 2010;11(1):68.

\bibitem{chakrabarti2012elf5}
Chakrabarti R, Hwang J, Blanco MA, Wei Y, Luka{\v{c}}i{\v{s}}in M, Romano RA,
  et~al.
\newblock Elf5 inhibits the epithelial--mesenchymal transition in mammary gland
  development and breast cancer metastasis by transcriptionally repressing
  Snail2.
\newblock Nature cell biology. 2012;14(11):1212--1222.

\bibitem{vernier2011regulation}
Vernier M, Bourdeau V, Gaumont-Leclerc MF, Moiseeva O, B{\'e}gin V, Saad F,
  et~al.
\newblock Regulation of E2Fs and senescence by PML nuclear bodies.
\newblock Genes \& development. 2011;25(1):41--50.

\bibitem{mcconnell1999induced}
McConnell BB, Gregory FJ, Stott FJ, Hara E, Peters G.
\newblock Induced expression of p16 INK4a inhibits both CDK4-and
  CDK2-associated kinase activity by reassortment of cyclin-CDK-inhibitor
  complexes.
\newblock Molecular and cellular biology. 1999;19(3):1981--1989.

\bibitem{villacanas2002structural}
Villaca{\~n}as {\'O}, P{\'e}rez JJ, Rubio-Mart{\'\i}nez J.
\newblock Structural analysis of the inhibition of Cdk4 and Cdk6 by p16INK4a
  through molecular dynamics simulations.
\newblock Journal of Biomolecular Structure and Dynamics. 2002;20(3):347--358.

\bibitem{bracken2007polycomb}
Bracken AP, Kleine-Kohlbrecher D, Dietrich N, Pasini D, Gargiulo G, Beekman C,
  et~al.
\newblock The Polycomb group proteins bind throughout the INK4A-ARF locus and
  are disassociated in senescent cells.
\newblock Genes \& development. 2007;21(5):525--530.

\bibitem{fang1999p21waf1}
Fang L, Igarashi M, Leung J, Sugrue MM, Lee SW, Aaronson SA.
\newblock p21Waf1/Cip1/Sdi1 induces permanent growth arrest with markers of
  replicative senescence in human tumor cells lacking functional p53.
\newblock Oncogene. 1999;18(18):2789--2797.

\bibitem{mao2012replicatively}
Mao Z, Ke Z, Gorbunova V, Seluanov A.
\newblock Replicatively senescent cells are arrested in G1 and G2 phases.
\newblock Aging (Albany NY). 2012;4(6):431.

\bibitem{chellappan1991e2f}
Chellappan SP, Hiebert S, Mudryj M, Horowitz JM, Nevins JR.
\newblock The E2F transcription factor is a cellular target for the RB protein.
\newblock Cell. 1991;65(6):1053--1061.

\bibitem{byeon1998tumor}
Byeon IJL, Li J, Ericson K, Selby TL, Tevelev A, Kim HJ, et~al.
\newblock Tumor Suppressor p16INK4A: Determination of Solution Structure and
  Analyses of Its Interaction with Cyclin-Dependent Kinase 4.
\newblock Molecular cell. 1998;1(3):421--431.

\bibitem{beausejour2003reversal}
Beaus{\'e}jour CM, Krtolica A, Galimi F, Narita M, Lowe SW, Yaswen P, et~al.
\newblock Reversal of human cellular senescence: roles of the p53 and p16
  pathways.
\newblock The EMBO journal. 2003;22(16):4212--4222.

\bibitem{freudlsperger2012tgf}
Freudlsperger C, Bian Y, Wise SC, Burnett J, Coupar J, Yang X, et~al.
\newblock TGF-$\beta$ and NF-$\kappa$B signal pathway cross-talk is mediated
  through TAK1 and SMAD7 in a subset of head and neck cancers.
\newblock Oncogene. 2012;32(12):1549--1559.

\bibitem{harley1990telomeres}
Harley C, Futcher A, Greider C.
\newblock Telomeres shorten during ageing of human fibroblasts.
\newblock Nature. 1990;345(6274):458--460.

\bibitem{mani2007mesenchyme}
Mani SA, Yang J, Brooks M, Schwaninger G, Zhou A, Miura N, et~al.
\newblock Mesenchyme Forkhead 1 (FOXC2) plays a key role in metastasis and is
  associated with aggressive basal-like breast cancers.
\newblock Proceedings of the National Academy of Sciences.
  2007;104(24):10069--10074.

\bibitem{zeisberg2009biomarkers}
Zeisberg M, Neilson EG, et~al.
\newblock Biomarkers for epithelial-mesenchymal transitions.
\newblock The Journal of clinical investigation. 2009;119(6):1429--1437.

\bibitem{bolos2003transcription}
Bol{\'o}s V, Peinado H, P{\'e}rez-Moreno MA, Fraga MF, Esteller M, Cano A.
\newblock The transcription factor Slug represses E-cadherin expression and
  induces epithelial to mesenchymal transitions: a comparison with Snail and
  E47 repressors.
\newblock Journal of cell science. 2003;116(3):499--511.

\bibitem{dave2011functional}
Dave N, Guaita-Esteruelas S, Gutarra S, Frias {\`A}, Beltran M, Peir{\'o} S,
  et~al.
\newblock Functional cooperation between Snail1 and twist in the regulation of
  ZEB1 expression during epithelial to mesenchymal transition.
\newblock Journal of Biological Chemistry. 2011;286(14):12024--12032.

\bibitem{casas2011snail2}
Casas E, Kim J, Bendesky A, Ohno-Machado L, Wolfe CJ, Yang J.
\newblock Snail2 is an essential mediator of Twist1-induced epithelial
  mesenchymal transition and metastasis.
\newblock Cancer research. 2011;71(1):245--254.

\bibitem{hajra2002slug}
Hajra KM, Chen DY, Fearon ER.
\newblock The SLUG zinc-finger protein represses E-cadherin in breast cancer.
\newblock Cancer research. 2002;62(6):1613--1618.

\bibitem{davila2015descriptive}
Davila-Velderrain J, Martinez-Garcia J, Alvarez-Buylla E.
\newblock Descriptive vs. Mechanistic Network Models in Plant Development in
  the Post-Genomic Era.
\newblock Plant Functional Genomics: Methods and Protocols. 2015;p. 455--479.

\bibitem{siegel2013cancer}
Siegel R, Naishadham D, Jemal A.
\newblock Cancer statistics, 2013.
\newblock CA: a cancer journal for clinicians. 2013;63(1):11--30.

\bibitem{jemal2011global}
Jemal A, Bray F, Center MM, Ferlay J, Ward E, Forman D.
\newblock Global cancer statistics.
\newblock CA: a cancer journal for clinicians. 2011;61(2):69--90.

\bibitem{stratton2009cancer}
Stratton MR, Campbell PJ, Futreal PA.
\newblock The cancer genome.
\newblock Nature. 2009;458(7239):719--724.

\bibitem{hudson2010international}
Hudson TJ, Anderson W, Aretz A, Barker AD, Bell C, Bernab{\'e} RR, et~al.
\newblock International network of cancer genome projects.
\newblock Nature. 2010;464(7291):993--998.

\bibitem{weinstein2013cancer}
Weinstein JN, Collisson EA, Mills GB, Shaw KRM, Ozenberger BA, Ellrott K,
  et~al.
\newblock The cancer genome atlas pan-cancer analysis project.
\newblock Nature genetics. 2013;45(10):1113--1120.

\bibitem{yaffe2013scientific}
Yaffe MB.
\newblock The scientific drunk and the lamppost: massive sequencing efforts in
  cancer discovery and treatment.
\newblock Science signaling. 2013;6(269):pe13.

\bibitem{creixell2012navigating}
Creixell P, Schoof EM, Erler JT, Linding R.
\newblock Navigating cancer network attractors for tumor-specific therapy.
\newblock Nature biotechnology. 2012;30(9):842--848.

\bibitem{depinho2004cancer}
DePinho RA, Polyak K.
\newblock Cancer chromosomes in crisis.
\newblock Nature genetics. 2004;36(9):932--934.

\bibitem{fujikawa2007ese}
Fujikawa M, Katagiri T, Tugores A, Nakamura Y, Ishikawa F.
\newblock ESE-3, an Ets family transcription factor, is up-regulated in
  cellular senescence.
\newblock Cancer science. 2007;98(9):1468--1475.

\bibitem{lee2006epithelial}
Lee JM, Dedhar S, Kalluri R, Thompson EW.
\newblock The epithelial--mesenchymal transition: new insights in signaling,
  development, and disease.
\newblock The Journal of cell biology. 2006;172(7):973--981.

\bibitem{cano2000transcription}
Cano A, P{\'e}rez-Moreno MA, Rodrigo I, Locascio A, Blanco MJ, del Barrio MG,
  et~al.
\newblock The transcription factor snail controls epithelial--mesenchymal
  transitions by repressing E-cadherin expression.
\newblock Nature cell biology. 2000;2(2):76--83.

\bibitem{sun2014slug}
Sun Y, Song GD, Sun N, Chen JQ, Yang SS.
\newblock Slug overexpression induces stemness and promotes hepatocellular
  carcinoma cell invasion and metastasis.
\newblock Oncology Letters. 2014;7(6):1936--1940.

\bibitem{liu2008zeb1}
Liu Y, El-Naggar S, Darling DS, Higashi Y, Dean DC.
\newblock Zeb1 links epithelial-mesenchymal transition and cellular senescence.
\newblock Development. 2008;135(3):579--588.

\bibitem{weinberg2008twisted}
Weinberg RA.
\newblock Twisted epithelial--mesenchymal transition blocks senescence.
\newblock Nature cell biology. 2008;10(9):1021--1023.

\bibitem{kim2006regulation}
Kim WY, Sharpless NE.
\newblock The Regulation of< i> INK4</i>/< i> ARF</i> in Cancer and Aging.
\newblock Cell. 2006;127(2):265--275.

\bibitem{mussel2010boolnet}
M{\"u}ssel C, Hopfensitz M, Kestler HA.
\newblock BoolNet---an R package for generation, reconstruction and analysis of
  Boolean networks.
\newblock Bioinformatics. 2010;26(10):1378--1380.

\bibitem{zhou2014relative}
Zhou JX, Samal A, d'H{\`e}rou{\"e}l AF, Price ND, Huang S.
\newblock Relative Stability of Network States in Boolean Network Models of
  Gene Regulation in Development.
\newblock arXiv preprint arXiv:14076117. 2014;.

\bibitem{wang2014quantitative}
Wang G, Zhu X, Gu J, Ao P.
\newblock Quantitative implementation of the endogenous molecular--cellular
  network hypothesis in hepatocellular carcinoma.
\newblock Interface focus. 2014;4(3):20130064.

\bibitem{zhu2015endogenous}
Zhu X, Yuan R, Hood L, Ao P.
\newblock Endogenous molecular-cellular hierarchical modeling of prostate
  carcinogenesis uncovers robust structure.
\newblock Progress in biophysics and molecular biology. 2015;.

\bibitem{li2013quantifying}
Li C, Wang J.
\newblock Quantifying cell fate decisions for differentiation and reprogramming
  of a human stem cell network: landscape and biological paths.
\newblock PLoS computational biology. 2013;9(8):e1003165.

\bibitem{zhou2011predicting}
Zhou JX, Brusch L, Huang S.
\newblock Predicting pancreas cell fate decisions and reprogramming with a
  hierarchical multi-attractor model.
\newblock PloS one. 2011;6(3):e14752.

\bibitem{Zhou2014Discrete}
Zhou JX, Qiu X, d'Herouel AF, Huang S.
\newblock Discrete Gene Network Models for Understanding Multicellularity and
  Cell Reprogramming: From Network Structure to Attractor Landscapes Landscape.
\newblock In: Computational Systems Biology Second Edition Elsevier. 2014;p.
  241--276.

\bibitem{lander2010edges}
Lander AD.
\newblock The edges of understanding.
\newblock BMC biology. 2010;8(1):40.

\bibitem{chen2005crucial}
Chen Z, Trotman LC, Shaffer D, Lin HK, Dotan ZA, Niki M, et~al.
\newblock Crucial role of p53-dependent cellular senescence in suppression of
  Pten-deficient tumorigenesis.
\newblock Nature. 2005;436(7051):725--730.

\bibitem{campisi2011cellular}
Campisi J.
\newblock Cellular senescence: putting the paradoxes in perspective.
\newblock Current opinion in genetics \& development. 2011;21(1):107--112.

\bibitem{narita2005senescence}
Narita M, Lowe SW.
\newblock Senescence comes of age.
\newblock Nature medicine. 2005;11(9):920--922.

\bibitem{yildiz2013genome}
Yildiz G, Arslan-Ergul A, Bagislar S, Konu O, Yuzugullu H, Gursoy-Yuzugullu O,
  et~al.
\newblock Genome-wide transcriptional reorganization associated with
  senescence-to-immortality switch during human hepatocellular carcinogenesis.
\newblock PloS one. 2013;8(5):e64016.

\bibitem{smit2010epithelial}
Smit MA, Peeper DS.
\newblock Epithelial-mesenchymal transition and senescence: two cancer-related
  processes are crossing paths.
\newblock Aging (Albany NY). 2010;2(10):735.

\bibitem{glaab2012enrichnet}
Glaab E, Baudot A, Krasnogor N, Schneider R, Valencia A.
\newblock EnrichNet: network-based gene set enrichment analysis.
\newblock Bioinformatics. 2012;28(18):i451--i457.

\bibitem{glaab2010topogsa}
Glaab E, Baudot A, Krasnogor N, Valencia A.
\newblock TopoGSA: network topological gene set analysis.
\newblock Bioinformatics. 2010;26(9):1271--1272.

\bibitem{winterhalter2014jepetto}
Winterhalter C, Widera P, Krasnogor N.
\newblock JEPETTO: a Cytoscape plugin for gene set enrichment and topological
  analysis based on interaction networks.
\newblock Bioinformatics. 2014;30(7):1029--1030.

\bibitem{saadatpour2013reduction}
Saadatpour A, Albert R, Reluga TC.
\newblock A reduction method for Boolean network models proven to conserve
  attractors.
\newblock SIAM Journal on Applied Dynamical Systems. 2013;12(4):1997--2011.

\bibitem{naldi2011dynamically}
Naldi A, Remy E, Thieffry D, Chaouiya C.
\newblock Dynamically consistent reduction of logical regulatory graphs.
\newblock Theoretical Computer Science. 2011;412(21):2207--2218.

\bibitem{allen2010introduction}
Allen LJ.
\newblock An introduction to stochastic processes with applications to biology.
\newblock CRC Press; 2010.

\bibitem{li2011regulatory}
Li J, Poi MJ, Tsai MD.
\newblock Regulatory mechanisms of tumor suppressor P16INK4A and their
  relevance to cancer.
\newblock Biochemistry. 2011;50(25):5566--5582.

\bibitem{yamakoshi2009real}
Yamakoshi K, Takahashi A, Hirota F, Nakayama R, Ishimaru N, Kubo Y, et~al.
\newblock Real-time in vivo imaging of p16Ink4a reveals cross talk with p53.
\newblock The Journal of cell biology. 2009;186(3):393--407.

\end{thebibliography}


\begin{figure}[p]
	\centering
	\includegraphics[width=150mm]{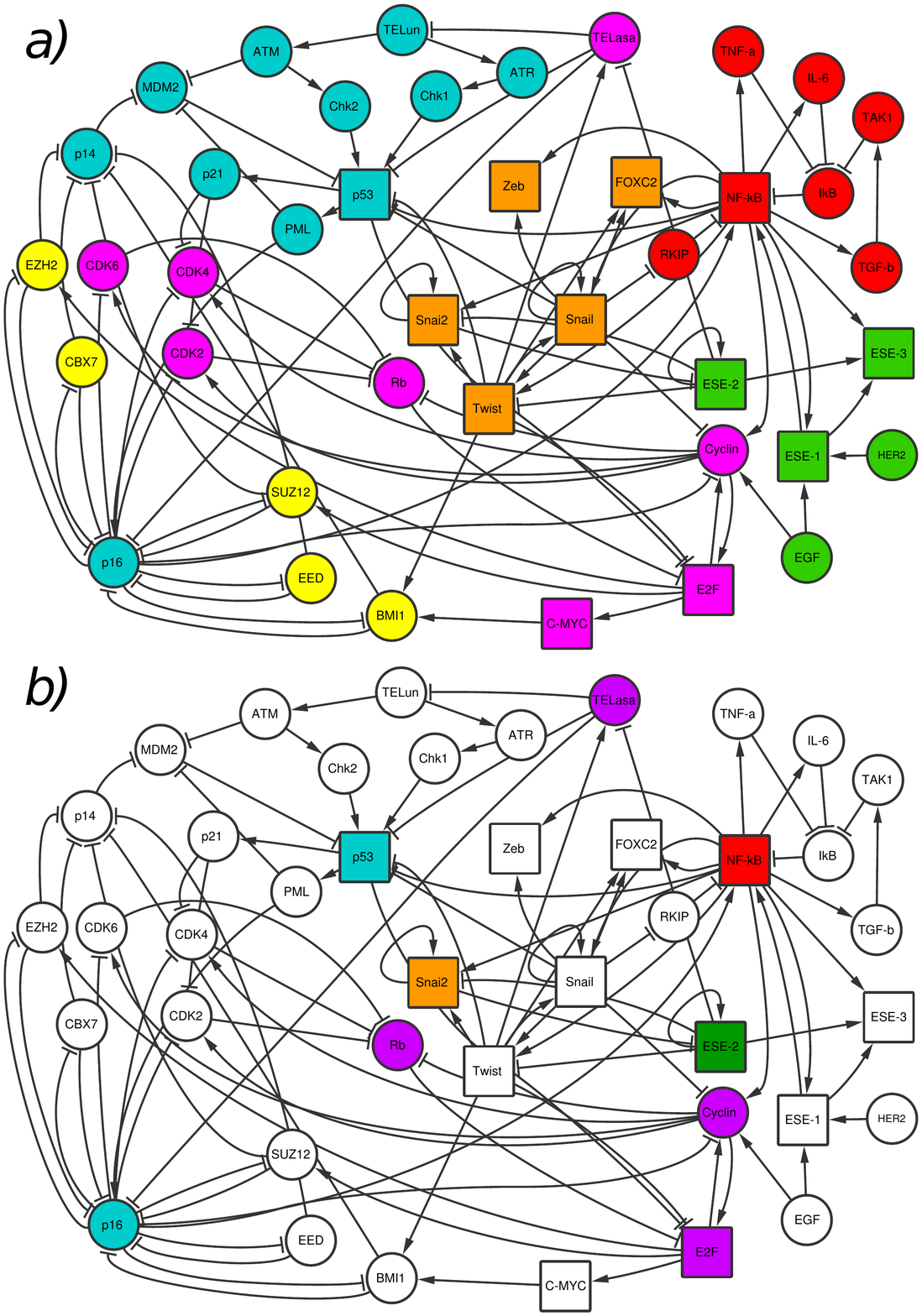}
	\caption{{\small Gene regulatory network for epithelial carcinogenesis.}}
	\label{fig:hgscores}
\end{figure}

\begin{figure}[p]
	\centering
	\includegraphics[width=150mm]{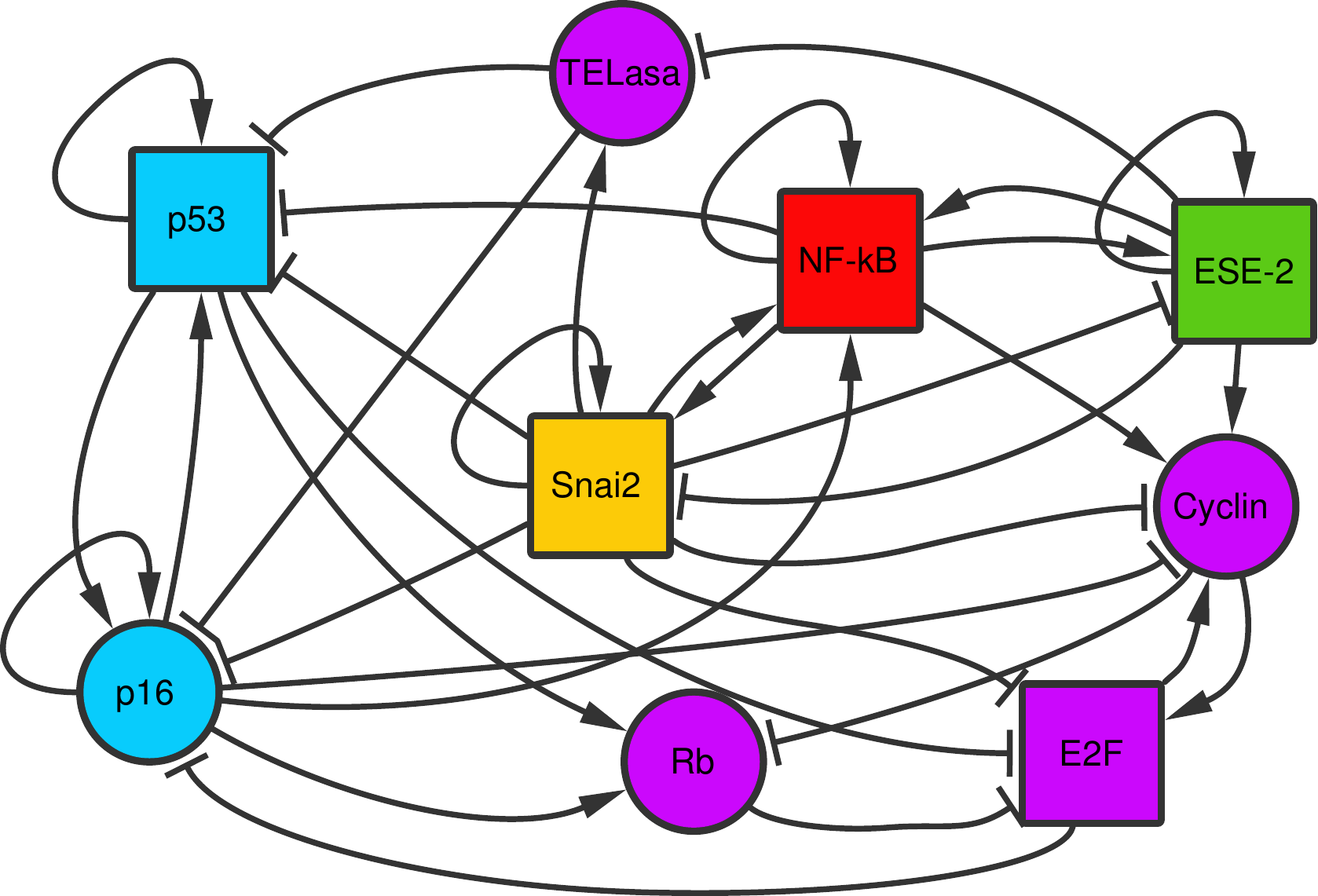}
	\caption{{\small Core gene regulatory network module for epithelial carcinogenesis.}}
	\label{fig:hgscores}
\end{figure}

\begin{figure}[p]
	\centering
	\includegraphics[width=150mm]{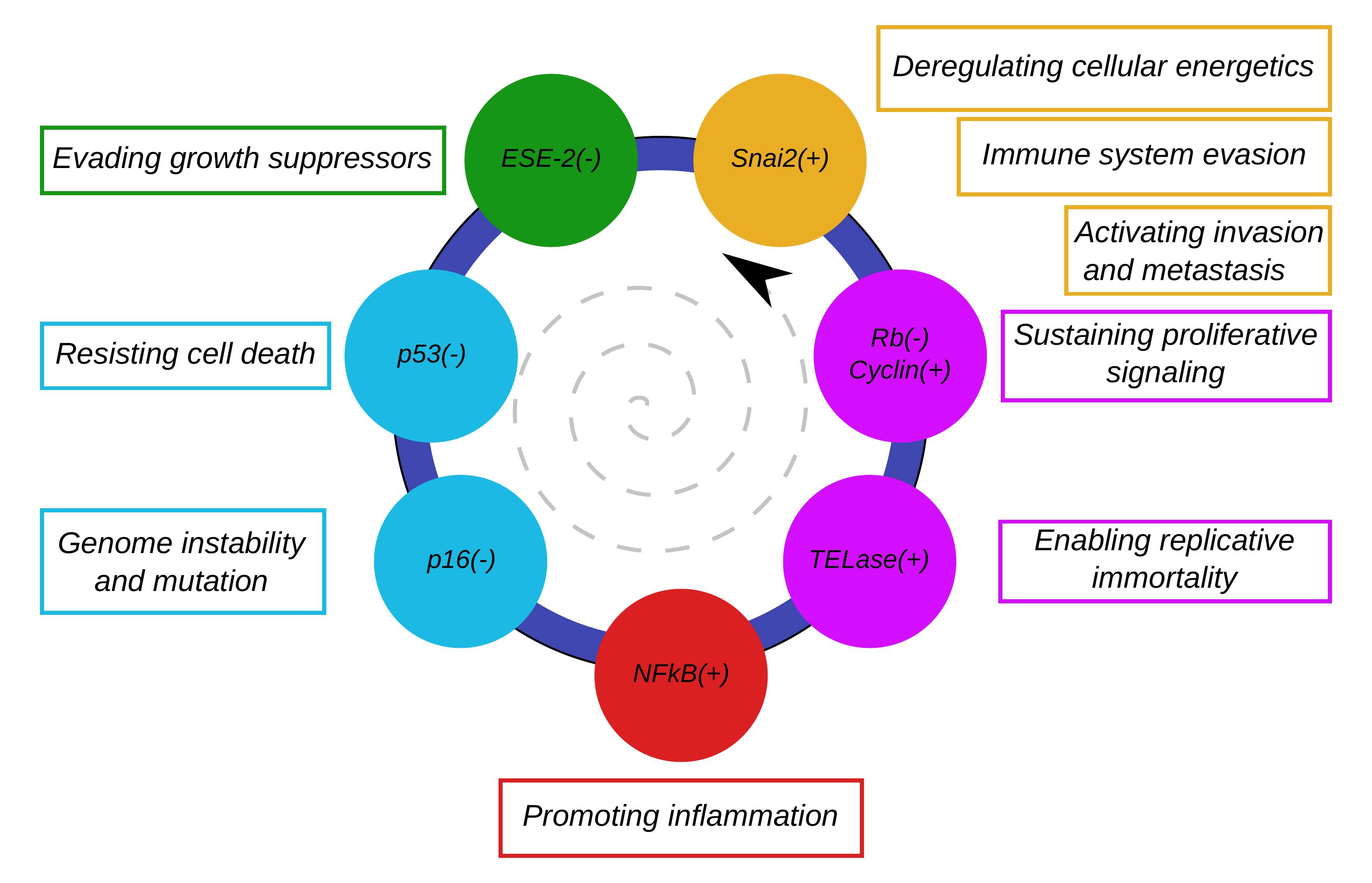}
	\caption{{\small The core gene regulatory module in the context of the {\em Hallmarks of Cancer} approach.}}
	\label{fig:hgscores}
\end{figure}

\begin{figure}[p]
	\centering
	\includegraphics[width=150mm]{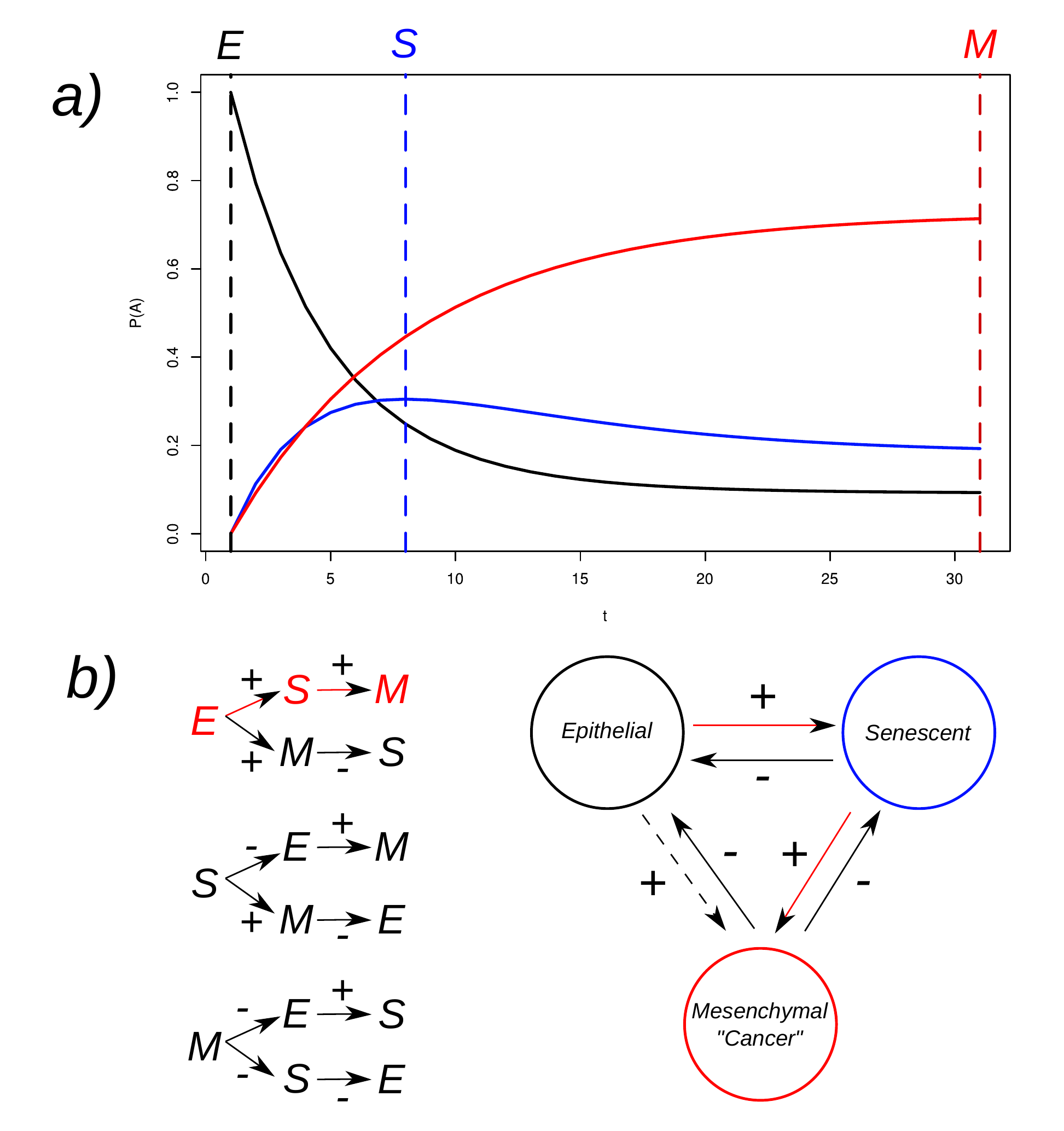}
	\caption{{\small Temporal sequence and global order of cell--fate attainment pattern under the stochastic Boolean GRN model during epithelial carcinogenesis.}}
	\label{fig:hgscores}
\end{figure}

\end{document}